\documentclass[11pt,a4paper]{article}
\usepackage[utf8]{inputenc}
\usepackage{csquotes}
\usepackage[T1]{fontenc}
\usepackage[american]{babel}
\usepackage{lmodern}
\usepackage[a4paper,margin=2.5cm]{geometry}

\usepackage{graphicx}
\usepackage{float}
\usepackage{amssymb}
\usepackage{amsmath}
\usepackage[
    style=phys,
    sorting=none,
    biblabel=brackets,
    citestyle=numeric-comp,
    maxbibnames=5,
    eprint=true,
    backend=biber
]{biblatex}

\usepackage[colorlinks=true, linkcolor=blue, citecolor=blue, urlcolor=blue]{hyperref}

\usepackage{authblk}
\usepackage{ragged2e}

\title{Massively Degenerate Coherent Perfect Absorption in Gradient-Index Fibers}

\author[1,*]{Helmut H{\"o}rner}
\author[2]{\c{S}ahin K.~{\"O}zdemir}
\author[1,*]{Stefan Rotter}

\affil[1]{Institute for Theoretical Physics, TU Wien, Wiedner Hauptstra{\ss}e 8-10/136, A-1040 Vienna, Austria}
\affil[2]{Department of Electrical and Computer Engineering, Saint Louis University, St. Louis, MO 63103, USA\linebreak}

\affil[*]{Corresponding authors:\linebreak \texttt{helmut.hoerner@tuwien.ac.at}, \texttt{stefan.rotter@tuwien.ac.at}}

\begin{document}

\maketitle

\begin{abstract}
Coherent perfect absorbers (CPAs) have recently attracted considerable attention due to their ability to enhance light--matter interaction. By exploiting interference, CPAs enable even weakly absorbing materials to achieve complete absorption under appropriate excitation conditions. Generalizing this concept to the simultaneous absorption of arbitrary multimode input states remains challenging, however, since conventional implementations typically operate only for a single or a very small number of input channels. Here, we propose a compact realization of a multimode coherent perfect absorber based on a gradient-index (GRIN) fiber. Using the self-imaging property of the fiber, the bulky free-space architecture of previous approaches is replaced by a monolithic waveguiding platform that supports near-degenerate rephasing of many spatial modes. We show that standard GRIN profiles optimized for minimal intermodal dispersion enable highly efficient absorption of complex multimode fields, with field-of-view reflectivities well below \(1\%\) for realistic parameters. This approach provides a practical and scalable route toward efficient multimode absorption in fiber-based and integrated photonic systems, with potential applications in light harvesting, optical control, and imaging.
\end{abstract}


\section{Introduction}

Absorbing light efficiently is a recurring objective in nature, physics, and engineering. While thick, strongly lossy media can extinguish light in a single pass, thin or weakly absorbing films are far less effective because the optical path through the absorbing material is short. A standard remedy is to place the absorber inside a resonator so that the optical field passes repeatedly through the absorbing region, thereby enhancing the effective loss \cite{Kishino1991,Uenlue1992}. When the external coupling to the cavity is balanced against the internal dissipation, the well-known condition of critical coupling occurs, at which perfect absorption is realized  \cite{Adler1969,Haus1984,Gorodetsky1999,Cai2000,yariv_2002}. 

The above strategy is particularly relevant when strong absorption is required in intrinsically thin or weakly absorbing elements. Examples include sensing layers, thin-film energy-harvesting structures, and compact optical control elements, where simply increasing the absorber thickness may be impractical or may compromise the intended functionality. Resonant enhancement therefore provides a way to achieve strong absorption while keeping the absorbing element physically small. Compact implementations are especially attractive for practical photonic platforms, where robustness, ease of alignment, and compatibility with fiber-based or integrated architectures are essential. Engineered waveguides also enable compact diffraction and refraction control \cite{Longhi2006}. Multimode fibers, in particular, are increasingly used as compact platforms for imaging, wavefront control, optical information transport, and complex spatial-mode manipulation \cite{Li2021,Kupianskyi2024,Mounaix2019,Valencia2020}.
In systems with multiple input channels or spatial modes, the concept of critical coupling generalizes to coherent perfect absorption (CPA), where perfect absorption is obtained for a particular set of relative amplitudes and phases across the inputs \cite{chong_coherent_2010,chong_hidden_2011,baranov_coherent_2017,Kishino1991,Uenlue1992,sweeney_perfectly_2019,Wang2021,Soleymani2022}. Since its introduction, CPA has been explored in a wide range of settings and for different functionalities, including complex and disordered structures \cite{Jiang2023,Sweeney2020,Dhia2018,Horodynski2022,pichler_random_2019,chen_perfect_2020}, single-port interferometric configurations \cite{Li2014,Jin2020}, optical switching \cite{Mock2012,Guo2023}, sensing \cite{Li2019,Zhang2022}, and all-optical logic elements \cite{EbrahimiMeymand2020,Fang2015}.

This interferometric selectivity of CPA is both powerful and restrictive. Only a few selected spatial states are perfectly absorbed, whereas inputs that do not match these states remain only partially absorbed. This tradeoff is evident across very different platforms, ranging from two-beam interferometric CPAs \cite{wan_time-reversed_2011,noh2012perfect,baranov_coherent_virtual_2017} to complex wavefronts interacting with disordered media \cite{pichler_random_2019}. Complementary approaches have explored how exceptional points, nonlinearities, and time-periodic scattering can be used to tailor the spectral and modal structure of CPA conditions \cite{Hoerner2024,Suwunnarat2022,Wang2024,Globosits2025}. A route to overcome the spatial-mode bottleneck was recently introduced through a massively degenerate coherent perfect absorber (MAD-CPA) \cite{Slobodkin2022}. The underlying idea follows directly from laser physics: formally, a CPA is the time-reversed counterpart of a laser at threshold \cite{noh2012perfect,longhi_pt-symmetric_2010}. It is therefore useful to seek the time reverse of a laser that can simultaneously oscillate in many spatial modes, since this suggests a way to absorb many spatial modes at once. Such devices are well known in the form of degenerate cavity lasers, which support a macroscopically large set of transverse modes with nearly identical eigenfrequencies \cite{arnaud1969degenerate,nixon2013observing,tradonsky2019}. In the MAD-CPA demonstrated in \cite{Slobodkin2022}, this idea is realized with a self-imaging free-space resonator containing a $4f$ telescope between two flat mirrors. Placing a weak absorber into this cavity even allows superpositions of arbitrary incident wavefronts, forming complex random speckle fields, to be absorbed with close-to-perfect efficiency through a massively parallel interference process.
While this free-space realization establishes an important proof of principle, it relies on several discrete optical elements and therefore results  in a  comparatively bulky and  alignment-sensitive setup, which is difficult to miniaturize. A more compact and robust implementation of MAD-CPA is therefore highly desirable.

The central requirement for transferring a MAD-CPA to a compact platform is to replace the self-imaging functionality of the free-space $4f$ cavity. A promising candidate for this role is a gradient-index (GRIN) fiber. In a GRIN fiber, the refractive index varies with radial position, which causes the transverse field to refocus periodically during propagation \cite{Gloge1973, Olshansky1976,Ghatak2004}. For suitable index profiles and fiber lengths, the input field is approximately reproduced after a characteristic propagation distance. A GRIN-fiber segment can therefore serve as a compact substitute for the two-lens $4f$ configuration of the original MAD-CPA. In this way, the essential physical ingredient of the concept -- the near-degenerate rephasing of many spatial modes at a common self-image plane -- can be realized in a monolithic geometry. Starting from this self-imaging principle, the following sections develop a compact MAD-CPA cavity based on a GRIN-fiber segment placed between a partially reflecting input coupler and a highly reflecting end mirror. The key question is how closely realistic GRIN profiles can reproduce the modal degeneracy required for massively degenerate coherent perfect absorption, and how this impacts the achievable absorption performance.

\section{Concept of the GRIN-Fiber MAD-CPA}

\subsection{Device Geometry}
\begin{figure}[H]
\centering
\includegraphics[width=0.75\linewidth]{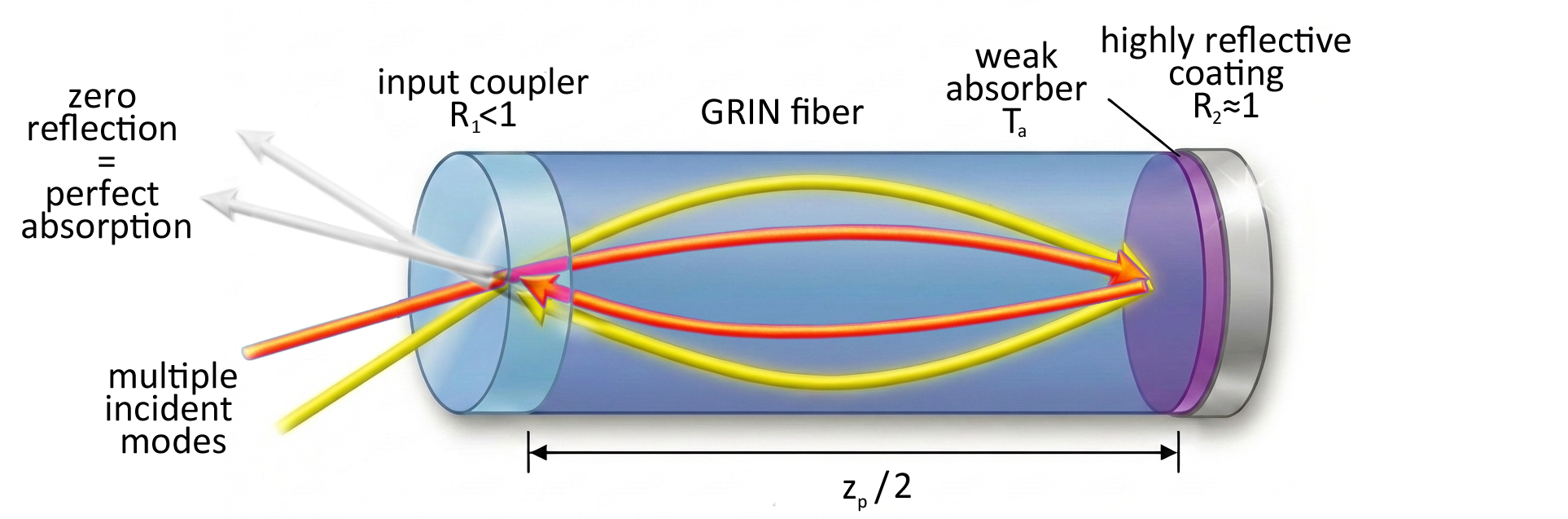}
\caption{Conceptual design of the GRIN-fiber MAD-CPA. A GRIN-fiber segment of length $L= z_p/2$ is bounded by a partially reflecting input coupler ($R_1<1$) and a highly reflecting end mirror ($R_2\approx 1$). Here \(z_p\) denotes the self-imaging pitch of the GRIN fiber, i.e.\ the propagation distance after which the transverse field approximately reproduces itself. A weak absorber with one-way power transmissivity $T_a$ provides the internal loss required for coherent perfect absorption. The colored arrows indicate two representative spatial modes that rephase on each cavity round trip due to the self-imaging condition, while the white arrows symbolize the suppressed reflected field under CPA operation.}
\label{fig:grin-cpa-sketch}
\end{figure}
Based on the self-imaging property of GRIN fibers introduced above, a compact massively degenerate coherent perfect absorber can be constructed by replacing the free-space $4f$ configration of Ref.~\cite{Slobodkin2022} with a GRIN-fiber segment. The resulting cavity geometry is sketched in Fig.~\ref{fig:grin-cpa-sketch}. A GRIN fiber of length $L = m\,z_p/2$ with $m\in\mathbb{N}$ is placed between a partially reflecting input coupler with power reflectivity $R_1<1$ and a highly reflecting end mirror with $R_2\approx 1$.  Here, \(z_p\) denotes the self-imaging pitch of the GRIN fiber, i.e. the propagation distance after which the transverse field reproduces itself. In a cavity of length \(L = m\,z_p/2\), the field is therefore re-imaged after each round trip, reproducing the central self-imaging functionality required for MAD-CPA operation. In contrast to the free-space realization, the self-imaging element is now contained in a compact and monolithic waveguiding structure, promising improved mechanical stability.

\subsection{CPA Matching Condition}

To obtain coherent perfect absorption, the cavity loss and the input coupling must satisfy the standard CPA matching condition \cite{Slobodkin2022}, which, for the case of an ideal end mirror ($R_2=1$), reduces to $R_1 = T_a^2$. Here, $T_a$ is the one-way \emph{power} transmissivity of the weak absorber, and $R_1$ is the power reflectivity of the input coupler. Under this condition, coherent perfect absorption occurs at those wavelengths for which the field directly reflected at the input coupler interferes destructively with the field returning from the cavity, causing the net reflected field to vanish. In the ideal self-imaging limit, this cancellation holds simultaneously for all supported spatial modes, so that the destructive interference required for CPA occurs for arbitrary superpositions of these modes. Note that the ray-optics picture of self-imaging provides an intuitive visualization, but the CPA condition is more naturally formulated in wave-optics terms. In that description, self-imaging corresponds to modal rephasing: the guided modes acquire the same round-trip phase modulo \(2\pi\). The cavity then acts as a massively degenerate CPA (MAD-CPA), capable in principle of absorbing arbitrary incident wavefronts that can be decomposed into the supported modal basis.

\subsection{Absorber Placement and Practical Realization}

The weak absorber may in principle be implemented as uniformly distributed loss in the fiber material. For a practical device, however, it is more attractive to realize it as a thin absorbing layer located at or near a self-image plane, for example between the right end of the GRIN fiber and the highly reflecting end mirror. In this configuration, the absorber is spatially well defined, its strength can be adjusted independently of the fiber design, and the GRIN segment itself can remain a low-loss transport element.

Likewise, the two cavity boundaries can be implemented in several straightforward ways. The input coupler can be realized by a partially reflective coating or coated interface at the left fiber facet, or alternatively by a fiber Bragg grating designed for the required partial reflectivity. The highly reflecting end may be formed by a dielectric high-reflectivity coating, a metallic mirror coating, or a bonded reflector placed directly after the absorber layer, or a high-reflectivity fiber Bragg grating. The key requirement is not the specific fabrication route, but the combination of self-imaging, weak absorption, and coupling-loss matching.

\section{Self-Imaging and Modal Rephasing in GRIN Fibers}

\subsection{Parabolic GRIN Fibers as the Standard Reference Case}

\begin{figure}[H]
\centering
\includegraphics[width=0.92\linewidth]{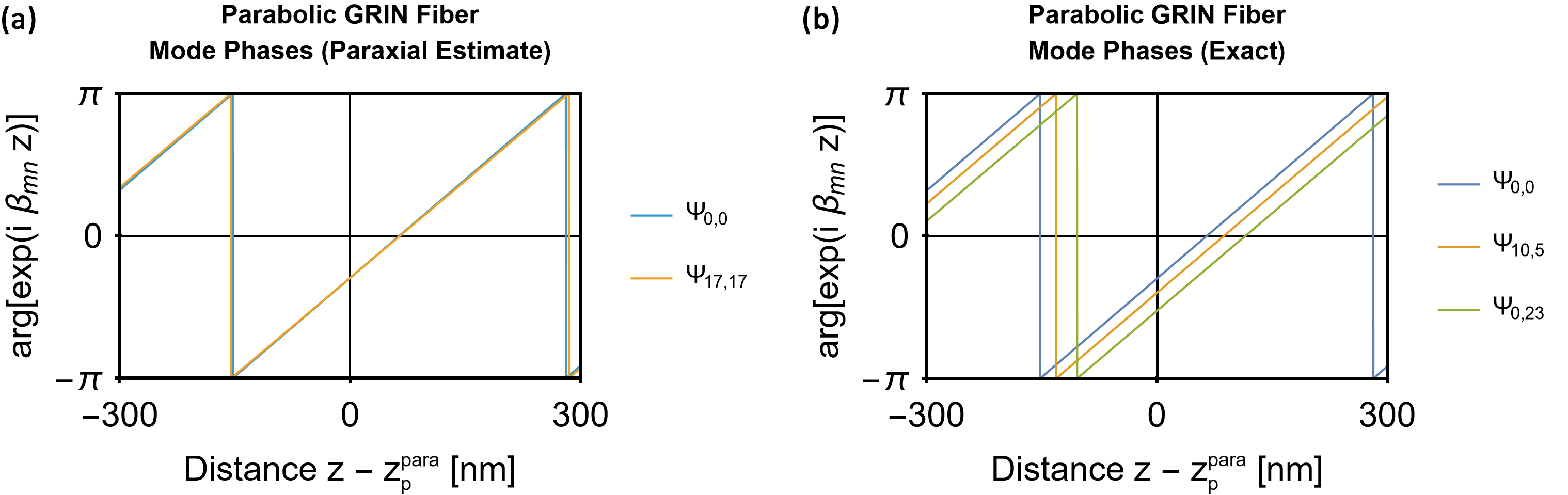}
\caption{Comparison of modal rephasing in a parabolic GRIN fiber with core-center refractive index $n_1=1.46$, relative index difference $\Delta=0.01$, and core-radius $a=25~\mu\mathrm{m}$, corresponding to a standard \(50/125~\mu\mathrm{m}\) graded-index multimode fiber, at $\lambda=633~\mathrm{nm}$. \textbf{(a)} Wrapped modal phases for the fundamental mode \(\psi_{0,0}\) and the high-order mode \(\psi_{17,17}\) in the paraxial approximation, showing exact phase coincidence at the paraxial pitch length \(z_p^{\mathrm{para}}\). These two modes are chosen to span a large range of mode orders while remaining within the estimated guided-mode set. \textbf{(b)} Wrapped modal phases for three representative modes computed from the exact propagation constants \(\beta_{mn}\) of Eq.~\eqref{eq:beta-mn-grin-paper}. At \(z=z_p^{\mathrm{para}}\), the phases no longer coincide exactly, revealing residual modal dephasing beyond the paraxial approximation. Only a narrow window of \(\pm 300~\mathrm{nm}\) around \(z_p^{\mathrm{para}}\) is shown, because for the present parameters \(z_p^{\mathrm{para}}\approx 1.1~\mathrm{mm}\). On the full propagation scale, the wrapped phases would exhibit so many \(2\pi\) jumps that the sawtooth curves would collapse visually into a dense block, obscuring the small but relevant mismatch near the expected self-imaging plane.
The actual optimum rephasing plane is shifted by about \(6.4~\mu\mathrm{m}\) from \(z_p^{\mathrm{para}}\) and is therefore well outside the plotted range; see Fig.~\ref{fig:z-opt-parabolic-paper}(a).}
\label{fig:mode-phases-parax-exact-paper}
\end{figure}

As a starting point, consider a GRIN fiber with the standard parabolic core profile \cite{Gloge1973, Olshansky1976,Ghatak2004}
\begin{equation}
n(r)=
\begin{cases}
n_1\,\sqrt{\,1-2\Delta\!\left(\dfrac{r}{a}\right)^{\!2}}, & 0\le r\le a,\\[6pt]
n_2, & r>a,
\end{cases}
\label{eq:parabolic_GRIN_paper}
\end{equation}
where $r$ is the radial coordinate from the fiber axis, $a$ is the core radius, $n_1$ and $n_2$ are the core-center and cladding indices, and
\begin{equation}
\Delta=\frac{n_1^2-n_2^2}{2n_1^2}\simeq \frac{n_1-n_2}{n_1}
\end{equation}
is the relative index difference between core and cladding.
This profile is the reference case because, in the paraxial picture, it leads to periodic self-imaging and thus provides the clearest connection to the free-space \(4f\) MAD-CPA. Within the weakly guiding approximation, the modal structure of the parabolic GRIN fiber is well described by Hermite-Gauss modes. Their propagation constants are \cite{Gloge1973, Olshansky1976,Ghatak2004}
\begin{equation}
\beta_{mn}
=
k_0 n_1\,
\sqrt{\,1
- \frac{2\,(m+n+1)}{k_0 n_1}\,
\sqrt{\frac{2\Delta}{a^{2}}}\,}\,,
\label{eq:beta-mn-grin-paper}
\end{equation}
with \(k_0=2\pi/\lambda\). In the paraxial limit, this expression reduces to
\begin{equation}
\beta_{mn}^{\mathrm{para}}
=
k_0 n_1 - (m+n+1)\frac{\sqrt{2\Delta}}{a},
\label{eq:beta-mn-grin-approx-paper}
\end{equation}
so that all modes rephase after the pitch length
\begin{equation}
z_p^{\mathrm{para}}=\frac{2\pi a}{\sqrt{2\Delta}}.
\label{eq:z_p_para_paper}
\end{equation}
This is the idealized self-imaging picture that motivates the use of GRIN fibers in the first place. Fig.~\ref{fig:mode-phases-parax-exact-paper}(a) shows the idealized paraxial picture: the wrapped phases of different modes intersect at \(z=z_p^{\mathrm{para}}\), indicating exact modal rephasing modulo \(2\pi\). If this behavior remained valid for all guided modes, a GRIN fiber would perfectly reproduce the fully degenerate self-imaging required for an ideal MAD-CPA. 

However, Fig.~\ref{fig:mode-phases-parax-exact-paper}(b) shows that this degeneracy is only approximate. Using the exact propagation constants from Eq.~\eqref{eq:beta-mn-grin-paper}, the modal phases no longer coincide perfectly at \(z_p^{\mathrm{para}}\), revealing residual modal dephasing beyond the paraxial approximation. The deeper reason is that, although the transverse mode equation of a parabolic GRIN fiber has the mathematical form of a harmonic-oscillator problem  \cite{Feit1979}, the equally spaced ladder in the full scalar Helmholtz equation appears in the quantity \(n_1^2k_0^2-\beta_{mn}^2\), not directly in the propagation constants \(\beta_{mn}\) themselves (see supplementary \ref{sec:supp-parabolic-grin-spectrum} for details). As a result, the exact \(\beta_{mn}\) follow the square-root dependence of Eq.~\eqref{eq:beta-mn-grin-paper} and are therefore no longer equally spaced in mode order. Only in the paraxial approximation of Eq.~\eqref{eq:beta-mn-grin-approx-paper}, where this square root is linearized, does one recover the regular spacing in \(\beta_{mn}\) required for exact periodic rephasing. This is important, because exact self-imaging requires that there exist a propagation distance \(z\) and a global phase \(\phi\) such that $e^{i\beta_{mn} z}=e^{i\phi}$
for all guided modes \((m,n)\). Equivalently, all modal phase differences must satisfy
\begin{equation}
(\beta_{mn}-\beta_{m'n'})z = 2\pi N_{mn,m'n'},
\qquad N_{mn,m'n'}\in\mathbb{Z},
\end{equation}
at one common propagation distance \(z\). In the paraxial approximation, this condition is fulfilled because the propagation constants depend linearly on the mode order \(m+n+1\), so the spacing between them is perfectly regular. As a result, all modal phase differences can line up simultaneously at \(z_p^{\mathrm{para}}\), as expressed by Eq.~\eqref{eq:z_p_para_paper}. Beyond the paraxial approximation, this regular spacing is lost: different pairs of modes accumulate phase differences at slightly different rates, so no single propagation distance can bring all modal phases into exact coincidence modulo \(2\pi\). One can therefore no longer expect exact self-imaging for an arbitrary superposition of guided modes, but only seek a plane of optimal phase clustering \cite{Bachmann1994,Soldano1995,Longhi2020}.

\begin{figure}[H]
\centering
\includegraphics[width=0.92\linewidth]{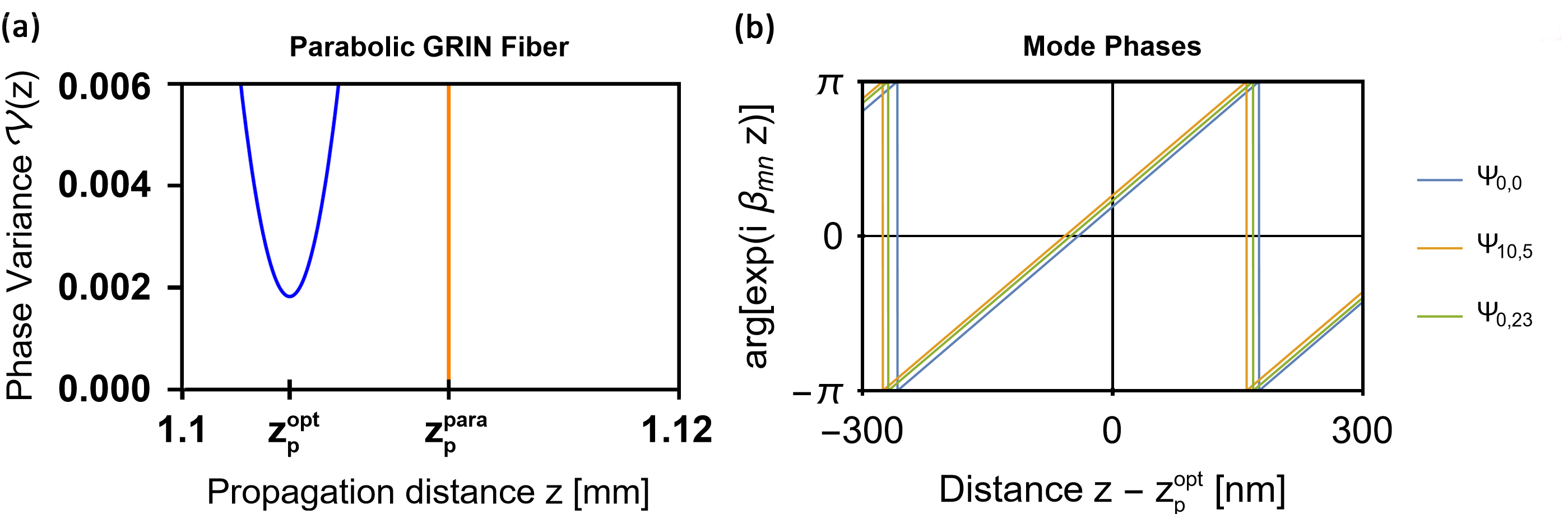}
\caption{Optimization of the self-imaging plane for a parabolic GRIN fiber (same parameters as in Fig.~\ref{fig:mode-phases-parax-exact-paper}). \textbf{(a)} Wrapped-phase variance $\mathcal{V}(z)$, evaluated over the first 300 spatial modes. The orange vertical line marks the paraxial estimate \(z_p^{\mathrm{para}} = 1.11072~\mathrm{mm}\), while the numerically optimized self-imaging plane \(z_p^{\mathrm{opt}} = 1.10432~\mathrm{mm}\) is shifted by about \(6.4~\mu\mathrm{m}\) from \(z_p^{\mathrm{para}}\). \textbf{(b)} Wrapped modal phases for the representative modes \(\psi_{0,0}\), \(\psi_{10,5}\), and \(\psi_{0,23}\), plotted versus the offset \(z-z_p^{\mathrm{opt}}\). These modes are chosen to span a broad range of mode orders within the truncated guided-mode set. Compared with the phase mismatch at \(z_p^{\mathrm{para}}\), shown in Fig.~\ref{fig:mode-phases-parax-exact-paper}(b), the phases now cluster much more closely at the optimized self-imaging plane.}
\label{fig:z-opt-parabolic-paper}
\end{figure}

But even in the absence of exact rephasing, one may still ask how closely the modal phases can be brought into alignment at a given propagation distance. To quantify the modal rephasing, a finite mode set
\begin{equation}
\mathcal{M}=\{(m,n):m,n\geq 0,\; m+n\leq M_{\max}\}
\end{equation}
is considered, and the wrapped phasors \(u_{mn}(z)=e^{i\beta_{mn}z}\) are evaluated. Rather than minimizing phase differences directly, we minimize the spread of the phasors \(u_{mn}(z)=e^{i\beta_{mn}z}\) in the complex plane. The reason is that the phase itself is defined only modulo \(2\pi\), so a direct variance of wrapped phases can be misleading: for example, two phases \(\phi_1=\varepsilon\) and \(\phi_2=2\pi-\varepsilon\) are physically almost identical, but their wrapped numerical values would appear to differ by nearly \(2\pi\). By working instead with the points \(u_{mn}(z)\) on the unit circle and minimizing the variances of their real and imaginary parts,
\begin{equation}
\mathcal{V}(z)=
\mathrm{Var}_{\mathcal{M}}\!\big[\Re\,u_{mn}(z)\big]
+
\mathrm{Var}_{\mathcal{M}}\!\big[\Im\,u_{mn}(z)\big],
\label{eq:GRIN-variance-paper}
\end{equation}
we measure the actual geometric clustering of the modal phasors on the unit circle. This is not literally the same as taking the variance of the wrapped phases, but it is the appropriate quantity for identifying phase alignment, because it treats phases that differ only by multiples of \(2\pi\) as close to one another.
The optimum self-imaging plane is then defined as
\begin{equation}
z_p^{\mathrm{opt}}
=
\arg\min_{|z-z_p^{\mathrm{para}}|\leq \delta}\,
\mathcal{V}(z),
\label{eq:zopt-variance-paper}
\end{equation}
with a small search window around the paraxial estimate. For representative parameters $n_1=1.46$, $\Delta=0.01$, $a=25~\mu\mathrm{m}$, and 
$\lambda=633~\mathrm{nm}$, Fig.~\ref{fig:z-opt-parabolic-paper}(a) shows that this procedure yields an optimized self-imaging distance  \(z_p^{\mathrm{opt}}\) slightly shifted from the paraxial value  \(z_p^{\mathrm{para}}\). For the numerical results shown in Fig.~\ref{fig:z-opt-parabolic-paper}(a), the functional \(\mathcal{V}(z)\) was evaluated in \textit{Wolfram Mathematica} 14.3 using the exact propagation constants from Eq.~\eqref{eq:beta-mn-grin-paper}. The mode set was truncated at \(M_{\max}=23\), corresponding to \(300\) Hermite-Gauss modes out of an estimated total of \(N\approx 656\) guided spatial modes for the representative fiber parameters stated above. The minimum in Eq.~\eqref{eq:zopt-variance-paper} is determined numerically in a search window of \(\pm 50~\mu\mathrm{m}\) around \(z_p^{\mathrm{para}}\) at the design wavelength \(\lambda=633~\mathrm{nm}\).\\

\newpage Choosing this optimized plane substantially improves the modal rephasing, as illustrated in Fig.~\ref{fig:z-opt-parabolic-paper}(b), where the wrapped phases of three representative modes cluster much more tightly around \(z=z_p^{\mathrm{opt}}\). The parabolic profile therefore already supports approximate self-imaging, but a significant residual phase spread remains. It is thus best viewed as the natural textbook starting point rather than the final design.

\subsection{Power-Law GRIN Profiles and the Optimum-Profile Condition}

\begin{figure}[H]
\centering
\includegraphics[width=0.92\linewidth]{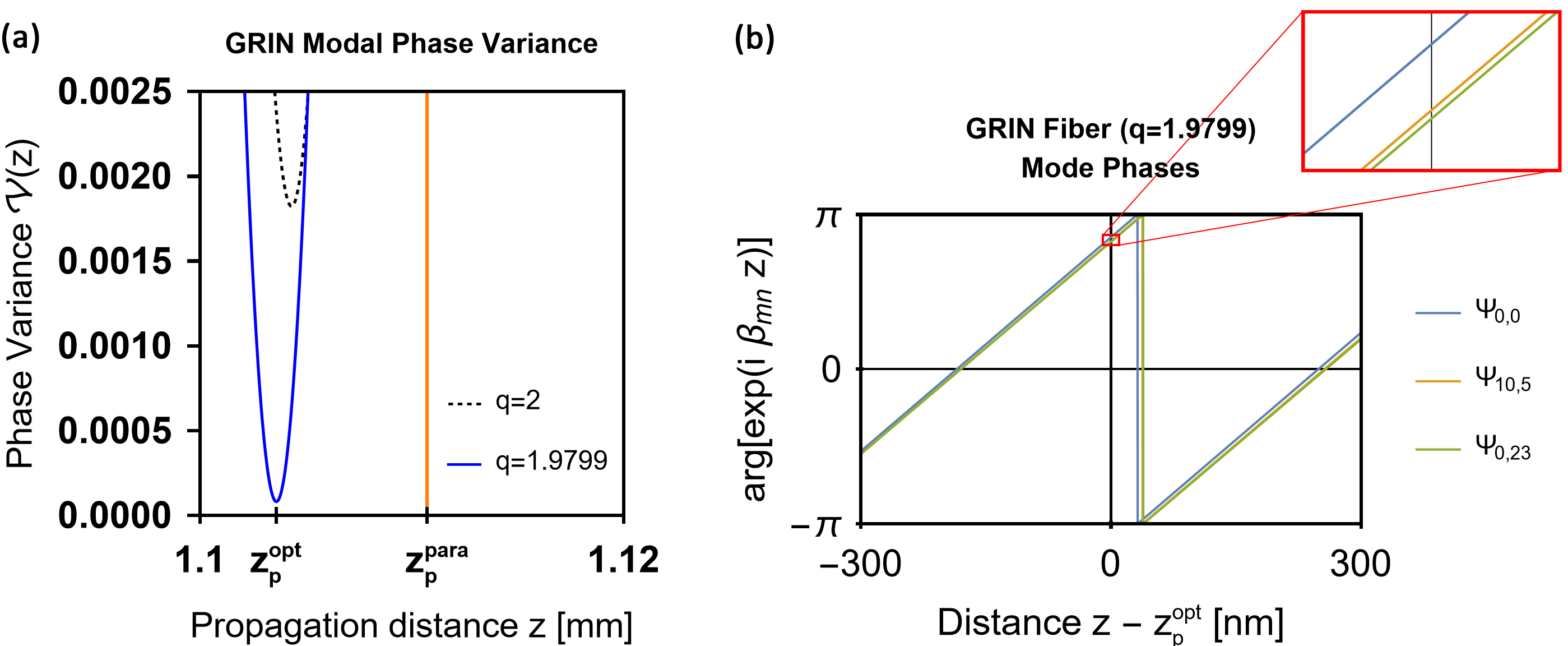}
\caption{Modal rephasing for power-law GRIN profiles (same parameters as in Fig.~\ref{fig:mode-phases-parax-exact-paper}). (a) Wrapped-phase variance \(\mathcal{V}(z)\) for the parabolic profile \(q=2\) and for the optimum-profile exponent \(q_0\) from Eq.~\eqref{eq:q0-opt-paper}, which minimizes intermodal group-delay dispersion. The optimum-profile case exhibits a substantially lower minimum, indicating improved modal rephasing. (b) Wrapped modal phases for the optimum-profile GRIN fiber, plotted versus the offset \(z-z_p^{\mathrm{opt}}\). The phases cluster more tightly than in the parabolic case, reflecting improved self-imaging and modal rephasing.}
\label{fig:z-opt-powerlaw-paper}
\end{figure}

For realistic multimode GRIN fibers designed to reduce intermodal group-delay dispersion in optical signal transport, the refractive-index profile is often modeled more generally as a power-law profile \cite{Gloge1973, Olshansky1976,Ghatak2004}
\begin{equation}
n(r)=
\begin{cases}
n_1\,\sqrt{\,1-2\Delta\!\left(\dfrac{r}{a}\right)^{\!q}}, & 0\le r\le a,\\[6pt]
n_2, & r>a,
\end{cases}
\label{eq:powerlaw-GRIN-paper}
\end{equation}
where the exponent \(q\) controls the radial grading. The parabolic case corresponds to $q=2$. In multimode GRIN fibers designed for optical signal transport, however, the exponent is commonly chosen slightly below 2 in order to minimize intermodal group-delay dispersion. For the power-law profile in Eq.~\eqref{eq:powerlaw-GRIN-paper}, the corresponding optimum exponent is \cite{Olshansky1976,Ghatak2004,Dakin2006a}
\begin{equation}
q_0 = 2\sqrt{\,1-2\Delta\,}.
\label{eq:q0-opt-paper}
\end{equation}
For the representative parameters $n_1=1.46$, $\Delta=0.01$, $a=25~\mu\mathrm{m}$, and $\lambda=633~\mathrm{nm}$, this evaluates to \(q_0=1.9799\). For a sufficiently large normalized frequency \(V=k_0 a n_1\sqrt{2\Delta}\gg 1\), i.e.\ in the strongly multimode regime, the propagation constants for the power-law profile are well approximated by the WKB expression \cite{Olshansky1976,Ghatak2004}.
\begin{equation}
\beta_{mn}
=
k_0 n_1\,
\sqrt{\,1
- 2\Delta\,
\left(
\frac{(m+n+1)^2}{\displaystyle \frac{q}{q+2}\,k_0^{2} a^{2} n_1^{2}\,\Delta}
\right)^{\!\frac{q}{q+2}} } \, .
\label{eq:beta-mn-wkb-powerlaw-paper}
\end{equation}
These propagation constants can be inserted into the same variance metric \(\mathcal{V}(z)\) defined in Eq.~\eqref{eq:GRIN-variance-paper}, allowing the modal rephasing properties of different profile exponents to be compared directly. Fig.~\ref{fig:z-opt-powerlaw-paper}(a) shows that the exponent \(q_0\), optimized for minimal intermodal group-delay dispersion, strongly improves the phase clustering compared with the parabolic profile. In other words, the profile used to reduce intermodal group-delay dispersion also improves single-wavelength modal rephasing at the self-imaging plane.

This is precisely the property required for a high-performance GRIN MAD-CPA. The corresponding phase alignment is illustrated in Fig.~\ref{fig:z-opt-powerlaw-paper}(b), where representative modes are plotted around the optimized self-imaging plane for the fiber with minimal intermodal group-delay dispersion. The clustering is visibly tighter than in the parabolic case.

\section{Multimode CPA Performance}

\subsection{Cavity Reflection Model}

To evaluate the GRIN-fiber MAD-CPA quantitatively, we treat each guided spatial mode independently and its complex reflection at the input port is calculated as follows: Let the GRIN segment have length \(L\), the partially reflecting input coupler be characterized by the complex-valued amplitude coefficients \(r_1\) and \(t_1\), and the end mirror by \(r_2\) and \(t_2\). A weak absorber with one-way \emph{amplitude} transmissivity \(t_a\) provides the internal loss required for CPA, so that the corresponding one-way \emph{power} transmissivity of the absorber is \(T_a=\vert t_a \vert ^2\). For a mode \((m,n)\) with propagation constant \(\beta_{mn}\) as given by Eq.~\eqref{eq:beta-mn-wkb-powerlaw-paper}, it is convenient to define the one-way complex propagation factor
\begin{equation}
p_{mn}=t_a\,e^{i\beta_{mn}L}.
\end{equation}
The corresponding round-trip factor is then
\begin{equation}
t_{\mathrm{RT}}^{(mn)} = r_2\,p_{mn}^2
= r_2\,t_a^2 e^{2i\beta_{mn}L}.
\label{eq:rt-factor-paper}
\end{equation}

The total reflected field at the input port is obtained by summing the directly reflected contribution and all cavity returns. The first term is the direct reflection \(r_1\), the second term corresponds to one cavity round trip, and the subsequent terms describe repeated round trips inside the cavity. This gives a geometric series in the round-trip factor \(r_1 t_{\mathrm{RT}}^{(mn)}\), so that the mode-resolved reflection coefficient can be written as
\begin{equation}
r_{mn}(L)
=
r_1 \;+\; t_1^{2}\,t_{\mathrm{RT}}^{(mn)}
     \sum_{k=0}^{\infty}\!\bigl(r_1 t_{\mathrm{RT}}^{(mn)}\bigr)^{k}
=
r_1+\frac{t_1^2\,t_{\mathrm{RT}}^{(mn)}}{1-r_1\,t_{\mathrm{RT}}^{(mn)}}
=
r_1+\frac{t_1^2\,r_2\,t_a^2\,e^{2i\beta_{mn}L}}
{1-r_1 r_2 t_a^2 e^{2i\beta_{mn}L}} .
\label{eq:r-mn-L-paper}
\end{equation}
\newpage
This expression already includes all multiple round trips inside the cavity. In the following, the case of a highly reflecting end mirror is considered, and the idealized limit \(R_2=1\) is used for the numerical evaluation presented below. The cavity length is chosen as $L=z_p^{\mathrm{opt}} / 2$ so that one round trip corresponds to propagation over the optimized self-imaging distance \(z_p^{\mathrm{opt}}\). The absorber strength and input coupling are chosen to satisfy the CPA matching condition discussed above. For \(R_2=1\), this condition reduces to $R_1=T_a^2$,
with \(R_1\) the power reflectivity of the input coupler and \(T_a\) the one-way power transmissivity of the absorber \cite{Slobodkin2022}.
Under these conditions, the mode-resolved reflection coefficient becomes
\begin{equation}
r_{mn}^{\mathrm{CPA}}
=
r_1+\frac{t_1^2\,r_2\,t_a^2\,e^{i\beta_{mn}z_p^{\mathrm{opt}}}}
{1-r_1 r_2 t_a^2 e^{i\beta_{mn}z_p^{\mathrm{opt}}}} .
\label{eq:r-mn-cpa-paper}
\end{equation}
In an ideal self-imaging cavity, all guided modes would satisfy the same round-trip phase condition modulo \(2\pi\), and Eq.~\eqref{eq:r-mn-cpa-paper} would vanish simultaneously for all of them at the CPA wavelength. In the GRIN-fiber case, residual modal dephasing prevents exact cancellation for every mode, but the optimized profiles discussed above can nevertheless make \(r_{mn}^{\mathrm{CPA}}\) very small over a large modal set. The performance of the device is therefore determined by how effectively the chosen GRIN profile suppresses the reflection of a highly multimode input field.

\subsection{Definition of the Multimode Input and Field-of-View Reflectivity}

To probe the GRIN-fiber MAD-CPA numerically under demanding multimode conditions, a spatially complex input field is constructed as an equal-weight superposition of many guided Hermite-Gauss modes with random relative phases. Specifically, let
\begin{equation}
\mathcal{M}_{N}=\{(m,n)\,:\, m,n\geq 0,\; m+n \leq M_{\max}\},
\end{equation}
with \(N=|\mathcal{M}_{N}|\) the number of included spatial modes. In the simulations below, the first \(300\) modes are used, out of an estimated total of \(N\approx 656\) guided spatial modes for the representative fiber parameters $n_1=1.46$, $\Delta=0.01$, $a=25~\mu\mathrm{m}$. The incident field is then written as
\begin{equation}
\psi_{\mathrm{in}}(x,y)
=
\mathcal{N}
\sum_{(m,n)\in\mathcal{M}_{N}}
e^{i\phi_{mn}}\,\psi_{mn}(x,y),
\label{eq:GRIN-psi-in-paper}
\end{equation}
where the phases \(\phi_{mn}\) are chosen independently and uniformly from \([0,2\pi)\), and \(\mathcal{N}\) is a real normalization factor. This choice represents a highly multimode input with no specially adapted wavefront engineering. It therefore provides a stringent test of the MAD-CPA concept: if the cavity suppresses the reflection of such a field, it is not merely operating for one selected mode, but for a broad and spatially complex modal superposition. The corresponding reflected field is obtained by weighting each mode in Eq.~\eqref{eq:GRIN-psi-in-paper} with its mode-resolved reflection coefficient \(r_{mn}^{\mathrm{CPA}}\) from Eq.~\eqref{eq:r-mn-cpa-paper},
\begin{equation}
\psi_{\mathrm{refl}}(x,y)
=
\mathcal{N}
\sum_{(m,n)\in\mathcal{M}_{N}}
r_{mn}^{\mathrm{CPA}}\,e^{i\phi_{mn}}\,\psi_{mn}(x,y).
\label{eq:psi-refl-paper}
\end{equation}
By construction, \(r_{mn}^{\mathrm{CPA}}\) already includes all multiple round trips inside the cavity, so Eq.~\eqref{eq:psi-refl-paper} is the total reflected field at the input port. An example of such a multimode input field is shown in Fig.~\ref{fig:in-out-q2-q0-paper}(a). The resulting spatial intensity distribution is strongly structured and speckle-like, as expected from the random-phase superposition of many guided modes. 
Within the scalar, weakly guiding approximation used here, the field-of-view reflectivity is defined from the longitudinal modal flux through the input plane. Since the transverse modes \(\psi_{mn}\) are normalized, the flux carried by each mode is proportional to \(\beta_{mn}\), and the equal-weight choice in Eq.~\eqref{eq:GRIN-psi-in-paper} implies \(|a_{mn}|=1\) for all modes. The field-of-view reflectivity therefore reduces to
\begin{equation}
R(\lambda)=
\frac{\sum_{(m,n)\in\mathcal{M}_{N}} \beta_{mn}(\lambda)\,\lvert r_{mn}(\lambda)\rvert^2}
{\sum_{(m,n)\in\mathcal{M}_{N}} \beta_{mn}(\lambda)}.
\label{eq:FOV-reflectivity-paper}
\end{equation}

\subsection{Performance of the Optimum-Profile GRIN MAD-CPA}

We now evaluate the GRIN MAD-CPA using the reflection model of Eq.~\eqref{eq:r-mn-cpa-paper} and the multimode input field defined in Eq.~\eqref{eq:GRIN-psi-in-paper}. For the numerical results in this section, the mode-resolved reflection coefficients \(r_{mn}(\lambda)\) are evaluated in \textit{Wolfram Mathematica} 14.3 from Eq.~\eqref{eq:r-mn-cpa-paper} for each mode in the truncated set \(\mathcal{M}\) with \(M_{\max}=23\), corresponding to \(300\) Hermite-Gauss modes. The same random multimode input field is used throughout the comparison, with equal modal amplitudes and fixed random phases (using a fixed random seed), so that the parabolic and dispersion-optimized profiles are tested under identical input conditions. For the parabolic case \(q=2\), the exact propagation constants from Eq.~\eqref{eq:beta-mn-grin-paper} are used together with the optimized self-imaging distance obtained from the phase-clustering procedure. For the dispersion-optimized profile, the exponent \(q_0\) from Eq.~\eqref{eq:q0-opt-paper} is inserted into the WKB propagation constants of Eq.~\eqref{eq:beta-mn-wkb-powerlaw-paper}, and the corresponding optimized self-imaging distance is determined in the same way.

\begin{figure}[H]
\centering
\includegraphics[width=0.82\linewidth]{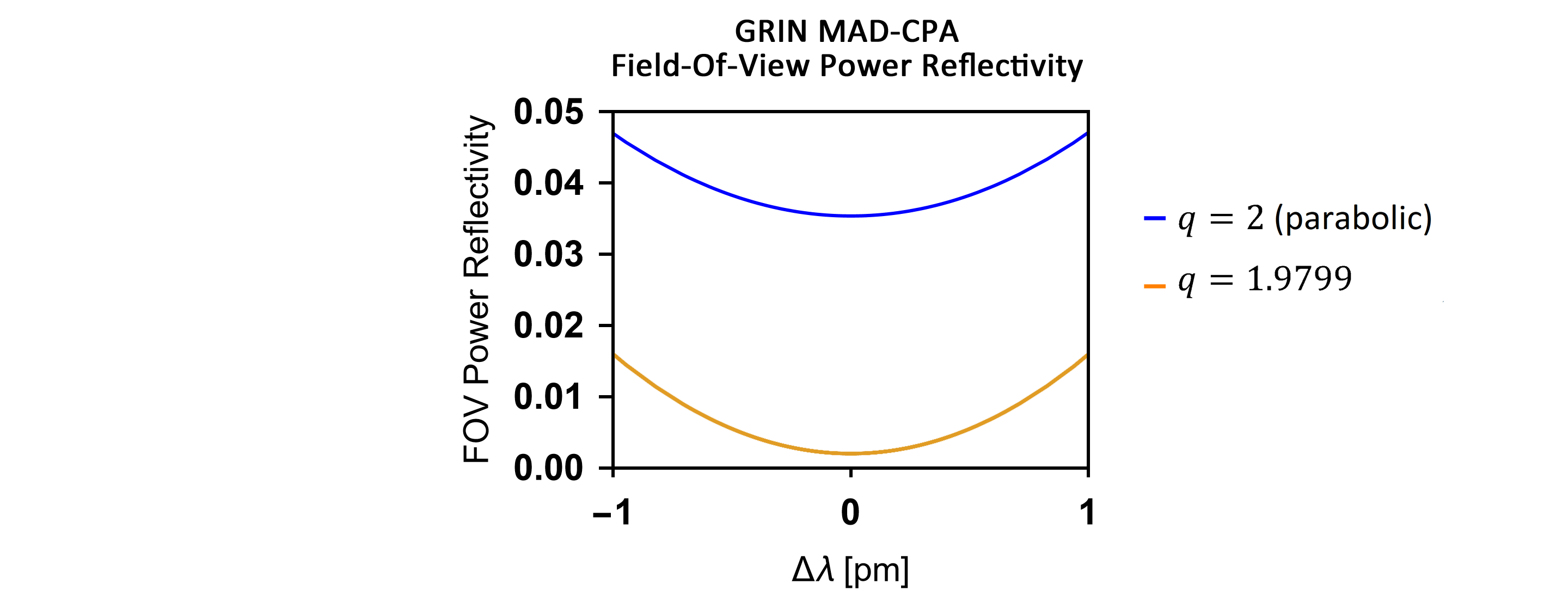}
\caption{Field-of-view power reflectivity \(R(\lambda)\) of the GRIN MAD-CPA around the resonance wavelength closest to \(\lambda=633~\mathrm{nm}\) for the representative GRIN-fiber parameters $n_1=1.46$, $\Delta=0.01$, and $a=25~\mu\mathrm{m}$, evaluated for the random equal-weight superposition of 300 guided Hermite-Gauss modes defined in Eq.~\eqref{eq:GRIN-psi-in-paper} and visualized in Fig.~\ref{fig:in-out-q2-q0-paper}(a). The parabolic profile \(q=2\) is compared with the profile optimized for minimal intermodal group-delay dispersion. For the profile condition minimizing inter-modal dispersion, \(q = 2\sqrt{1-2\Delta}\), the same fiber parameters give \(q=1.9799\). This profile design exhibits a substantially deeper minimum, demonstrating strongly improved multimode absorption.}
\label{fig:refl-spectrum-paper}
\end{figure}

\begin{figure}[H]
\centering
\includegraphics[width=0.98\linewidth]{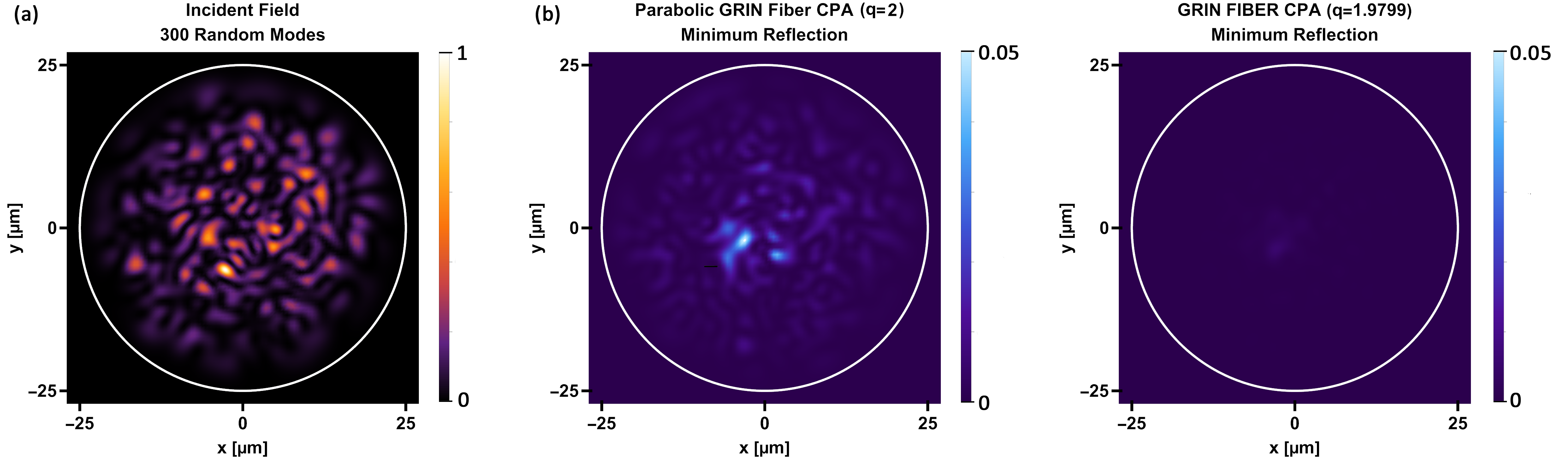}
\caption{Incident and reflected field intensities for a highly multimode input in the GRIN MAD-CPA (same parameters as in Fig.~\ref{fig:refl-spectrum-paper}). \textbf{(a)}~Incident field intensity \(|\psi_{\mathrm{in}}(x,y)|^2\) formed by an equal-weight superposition of 300 Hermite-Gauss modes with random phases. \textbf{(b)}~Reflected intensity \(|\psi_{\mathrm{refl}}(x,y)|^2\) at the wavelength of minimum FOV reflectivity for the parabolic GRIN MAD-CPA (\(q=2\)). \textbf{(c)}~Reflected intensity at the wavelength of minimum FOV reflectivity for the optimum-profile GRIN MAD-CPA with exponent \(q_0\) from Eq.~\eqref{eq:q0-opt-paper}. In all panels, the white circle marks the GRIN-fiber core boundary at radius \(a\). Notice the different color bar in panel (b) and (c) as compared to panel (a).}
\label{fig:in-out-q2-q0-paper}
\end{figure}

The reflected field is then constructed as the coherent superposition of all reflected modes,
\[
\psi_{\mathrm{refl}}(x,y;\lambda)=\sum_{(m,n)\in\mathcal{M}} r_{mn}(\lambda)\,a_{mn}\,\psi_{mn}(x,y),
\]
with the same input coefficients \(a_{mn}\) as in Eq.~\eqref{eq:GRIN-psi-in-paper}. The field-of-view reflectivity \(R(\lambda)\) is obtained by numerical evaluation of Eq.~\eqref{eq:FOV-reflectivity-paper}, and the wavelength of minimum reflectivity is determined numerically. The spectra shown in Fig.~\ref{fig:refl-spectrum-paper} are generated by sampling \(R(\lambda)\) in a narrow wavelength interval around this minimum and interpolating the resulting data points. The reflected-field intensity maps in Fig.~\ref{fig:in-out-q2-q0-paper} are evaluated at the respective minimizing wavelengths and normalized to the peak intensity of the incident field.

As a baseline, we first consider the textbook parabolic GRIN profile with \(q=2\), using the exact propagation constants from Eq.~\eqref{eq:beta-mn-grin-paper} and the optimized self-imaging distance \(z_p^{\mathrm{opt}}\) obtained from the phase-clustering procedure above. We then compare this case with the optimum-profile GRIN fiber, whose exponent \(q_0\) is given by Eq.~\eqref{eq:q0-opt-paper} and whose propagation constants are calculated from the WKB expression in Eq.~\eqref{eq:beta-mn-wkb-powerlaw-paper}. Fig.~\ref{fig:refl-spectrum-paper} shows the corresponding field-of-view (FOV) power reflectivity \(R(\lambda)\) for the two profiles around the resonance wavelength, evaluated for the representative GRIN-fiber parameters $n_1=1.46$, $\Delta=0.01$, and $a=25~\mu\mathrm{m}$.

For our representative fiber parameters, the parabolic profile yields a minimum field-of-view power reflectivity of approximately \(R_{\min}\approx 3.5\%\), whereas the optimum-profile design reduces this value to approximately \(R_{\min}\approx 0.2\%\). The improvement is a direct consequence of the tighter modal rephasing established in the previous section: because the guided modes accumulate nearly the same round-trip phase at the optimized self-imaging plane, the CPA condition is satisfied much more uniformly across the modal set.  

The corresponding spatial intensity patterns are shown in Fig.~\ref{fig:in-out-q2-q0-paper}. The random-speckle incident field shown in Fig.~\ref{fig:in-out-q2-q0-paper}(a) is a superposition of 300 guided modes with the same amplitude and random phases. At the wavelength of minimum reflectivity, the parabolic GRIN MAD-CPA still leaves a clearly visible residual reflected field distributed over a sizeable part of the core, as shown in Fig.~\ref{fig:in-out-q2-q0-paper}(b). 

In contrast, Fig.~\ref{fig:in-out-q2-q0-paper}(c) shows that the optimum-profile design suppresses the reflected field much more efficiently, leaving only a weak residual feature near the center. Note that the color scales in the two reflected-field panels (b) and (c) differ by one order of magnitude. Thus, the dominant performance gain is obtained by moving from the textbook parabolic GRIN profile to the standard optimum-profile design known from multimode-fiber theory. This is an important practical result: GRIN fibers with near-parabolic optimum profiles, as used in multimode fiber design for telecommunication purposes, are well suited to transfer the MAD-CPA concept from a bulky free-space proof-of-principle system to a compact waveguiding platform with very high multimode absorption.

\subsection{Incremental Improvement from Numerically Refined Profiles}

\begin{figure}[H]
\centering
\includegraphics[width=0.98\linewidth]{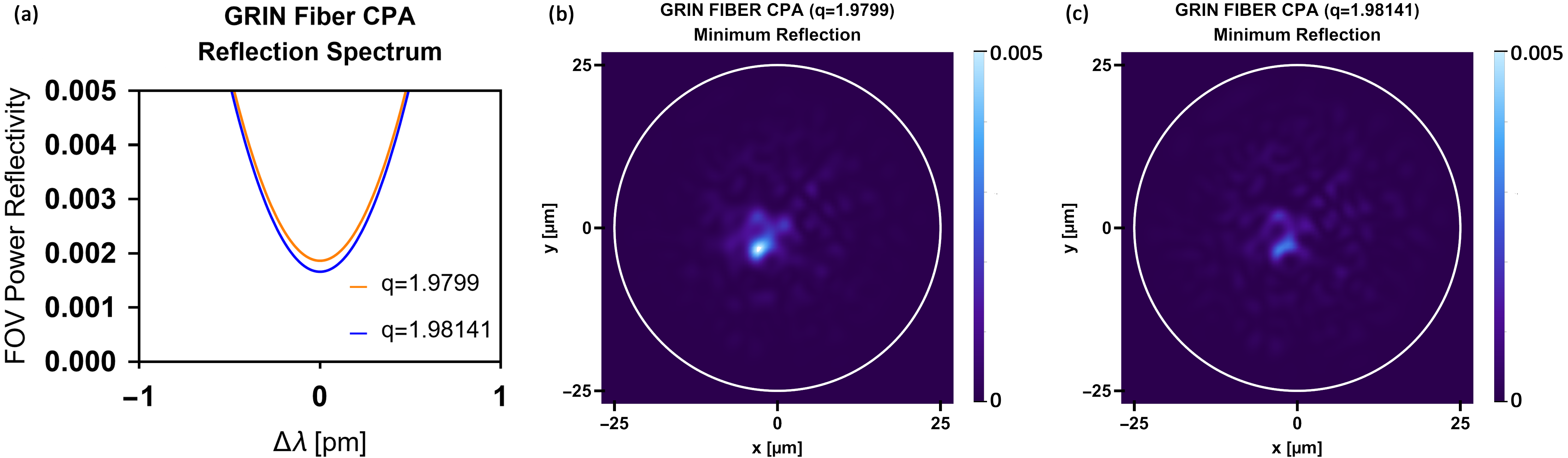}
\caption{Incremental improvement from numerical refinement of the GRIN profile exponent. \textbf{(a)} FOV power reflectivity around resonance for the optimum-profile case \(q_0=1.9799\) and the numerically refined value \(q=1.98141\). \textbf{(b,c)} Reflected-field intensity at the wavelength of minimum reflectivity for the two cases. The additional improvement is visible but modest.}
\label{fig:incremental-improvement-paper}
\end{figure}

A numerical refinement of the profile exponent \(q\) can reduce the residual reflectivity slightly further, but only by a small amount. To determine the numerically refined exponent, the phase-clustering functional \(\mathcal{V}(z)\) is evaluated with \textit{Wolfram Mathematica} 14.3 for a dense set of trial exponents \(q\) in a narrow interval around \(q_0\). For each trial value, the corresponding optimum self-imaging plane \(z_p^{\mathrm{opt}}(q)\) is first obtained from Eq.~\eqref{eq:zopt-variance-paper}, and the minimum value \(\mathcal{V}_{\min}(q)\) is recorded. The resulting dependence of \(\mathcal{V}_{\min}\) on \(q\) is then interpolated and minimized numerically, which yields the refined value \(q=1.98141\). As Fig.~\ref{fig:incremental-improvement-paper} shows, this additional optimization produces only a modest improvement over the exponent \(q_0\) obtained from the standard condition for minimal intermodal group-delay dispersion. The reason for this is that optimizing the exponent in Eq.~\eqref{eq:beta-mn-wkb-powerlaw-paper} for minimal intermodal dispersion and optimizing it for maximal modal rephasing at a fixed wavelength lead, to first order, to the same result, and in second order to almost the same result; see Section~\ref{sec:supp-rephasing-vs-dmd} of the Supplementary Material for details.

This is an important practical conclusion: the strongest improvement over the parabolic textbook case is already obtained from the standard optimum-profile design, without requiring substantial additional tailoring of the radial index profile. Commercial graded-index multimode fibers optimized for low modal dispersion are widely available, for example in the form of standard OM2/OM3/OM4/OM5 fibers.

\section{Conclusion}
A compact implementation of massively degenerate coherent perfect absorption has been proposed based on a gradient-index (GRIN) fiber. The central idea is to replace the bulky free-space \(4f\) self-imaging stage of the original MAD-CPA concept \cite{Slobodkin2022} by the intrinsic self-imaging property of a GRIN-fiber segment. In this way, the essential physical mechanism of degenerate coherent perfect absorption is transferred to a monolithic waveguiding platform. Our analysis shows that a strictly parabolic GRIN profile provides the natural textbook starting point, but that GRIN fibers optimized for minimum intermodal dispersion, as commonly used in telecommunications, are considerably better suited for the present purpose. Although realistic cylindrical GRIN fibers do not exhibit exact modal degeneracy, profiles optimized for low intermodal dispersion -- including commercially available graded-index multimode fibers -- lead to strongly improved modal rephasing at a fixed wavelength and therefore to much deeper multimode coherent perfect absorption. For the representative parameters considered here, the minimum field-of-view reflectivity decreases from approximately \(3.5\%\) for the parabolic case to approximately \(0.2\%\) for the optimum-profile design, while a further numerical refinement of the profile exponent yields only a small additional improvement. These results indicate that GRIN fibers with near-parabolic optimum profiles, as used in multimode fiber design for telecommunication purposes, are well suited to transfer the MAD-CPA concept from a bulky free-space proof-of-principle system to a compact waveguiding platform with very high multimode absorption. This opens a promising path toward practical implementations of massively degenerate coherent perfect absorbers for applications in fiber photonics, optical control, light harvesting, and imaging.

\medskip

\printbibliography[heading=bibintoc, title={References}]

\clearpage
\appendix

\begin{center}
{\LARGE\bfseries Supplementary Material for\\[0.5em]
Massively Degenerate Coherent Perfect Absorption in Gradient-Index Fibers\par}

\vspace{1.5em}

{\large Helmut H{\"o}rner, {\c{S}}ahin K.~{\"O}zdemir, and Stefan Rotter\par}
\end{center}

\vspace{2em}

\renewcommand{\thesection}{S\arabic{section}}
\renewcommand{\thesubsection}{S\arabic{section}.\arabic{subsection}}
\renewcommand{\theequation}{S\arabic{section}.\arabic{equation}}
\renewcommand{\thefigure}{S\arabic{section}.\arabic{figure}}

\numberwithin{equation}{section}
\numberwithin{figure}{section}

\addcontentsline{toc}{section}{Supplementary Material}

\section{Why a Parabolic GRIN Profile Does Not Produce Exact Rephasing Beyond the Paraxial Approximation}
\label{sec:supp-parabolic-grin-spectrum}

This supplementary section explains why a parabolic GRIN profile leads to exact modal rephasing only in the paraxial approximation, while the full scalar Helmholtz equation yields a residual phase spread. The central point is subtle but important: for a parabolic profile, the exact transverse mode equation is indeed mathematically equivalent to a harmonic-oscillator eigenvalue problem \cite{Feit1979}, but the equally spaced ladder appears in the quantity \(n_1^2 k_0^2 - \beta^2\), not directly in the propagation constants \(\beta\) themselves. Only after a paraxial linearization does this become an approximately equally spaced \(\beta\)-spectrum, which is the prerequisite for exact global intermodal rephasing.

\subsection{Starting Point: The Scalar Helmholtz Equation}

For a monochromatic scalar field \(E(x,z)\) with vacuum wavenumber \(k_0 = 2\pi/\lambda\), the scalar Helmholtz equation reads
\begin{equation}
\frac{\partial^2 E}{\partial x^2}
+
\frac{\partial^2 E}{\partial z^2}
+
k_0^2 n^2(x)\,E
= 0.
\label{eq:supp-helmholtz-1d}
\end{equation}
To keep the derivation transparent, we first use a single transverse coordinate \(x\). The corresponding two-dimensional transverse problem in \(x\) and \(y\) is a straightforward extension and is discussed below.

\subsection{Paraxial Approximation} 

We first recall the standard paraxial derivation, because this is the origin of the familiar harmonic-oscillator picture. 

\subsubsection*{Slowly Varying Envelope Ansatz} 

Assume that the field propagates mainly in the positive \(z\)-direction and write 

\begin{equation} E(x,z) = A(x,z)\,e^{i n_0 k_0 z}, \label{eq:supp-envelope-ansatz} 
\end{equation} 

where \(n_0\) is a reference index and \(A(x,z)\) is a slowly varying envelope. Inserting Eq.~\eqref{eq:supp-envelope-ansatz} into Eq.~\eqref{eq:supp-helmholtz-1d} gives 

\begin{equation} \frac{\partial^2 A}{\partial x^2} + \frac{\partial^2 A}{\partial z^2} + 2 i n_0 k_0 \frac{\partial A}{\partial z} + k_0^2 \bigl(n^2(x)-n_0^2\bigr) A = 0. \label{eq:supp-envelope-exact} \end{equation} 

The paraxial approximation neglects the second derivative \(\partial^2 A/\partial z^2\) compared with the first derivative term \(2 i n_0 k_0 \, \partial A/\partial z\). One then obtains 

\begin{equation} i\frac{\partial A}{\partial z} = -\frac{1}{2 n_0 k_0}\frac{\partial^2 A}{\partial x^2} -\frac{k_0}{2 n_0}\bigl(n^2(x)-n_0^2\bigr)A. \label{eq:supp-paraxial-wave} 
\end{equation} 

\subsubsection*{Parabolic Profile}

Now choose a parabolic refractive-index profile in the form 

\begin{equation} n^2(x) = n_0^2 - \alpha x^2, \label{eq:supp-parabolic-profile-1d} 
\end{equation} 

with \(\alpha > 0\). Equation~\eqref{eq:supp-paraxial-wave} then becomes 

\begin{equation} i\frac{\partial A}{\partial z} = -\frac{1}{2 n_0 k_0}\frac{\partial^2 A}{\partial x^2} + \frac{k_0 \alpha}{2 n_0}\,x^2 A. \label{eq:supp-paraxial-ho} 
\end{equation} This has exactly the form of a Schrödinger equation for a harmonic oscillator, with \(z\) playing the role of time. We therefore seek modal solutions of the form 

\begin{equation} A_n(x,z) = \psi_n(x)\,e^{-i \mu_n z}. \label{eq:supp-paraxial-mode-ansatz} 
\end{equation} 

Inserting Eq.~\eqref{eq:supp-paraxial-mode-ansatz} into Eq.~\eqref{eq:supp-paraxial-ho} gives 

\begin{equation} \mu_n \psi_n = -\frac{1}{2 n_0 k_0}\frac{d^2\psi_n}{dx^2} + \frac{k_0 \alpha}{2 n_0}\,x^2 \psi_n, \label{eq:supp-paraxial-eigenproblem} 
\end{equation} 

which can be rearranged as 

\begin{equation} -\frac{d^2\psi_n}{dx^2} + \alpha k_0^2 x^2 \psi_n = 2 n_0 k_0 \mu_n \,\psi_n. \label{eq:supp-paraxial-ho-rearranged} 
\end{equation} 

This is the standard harmonic-oscillator eigenvalue problem, whose eigenvalues are equally spaced: 

\begin{equation} \mu_n = \Omega \left(n+\frac12\right), \qquad n=0,1,2,\dots \label{eq:supp-paraxial-mu} 
\end{equation} 

for some constant \(\Omega\) determined by the coefficients in Eq.~\eqref{eq:supp-paraxial-eigenproblem}. Returning to the full field \(E\), one has 

\begin{equation} E_n(x,z) = \psi_n(x)\,e^{i(n_0 k_0 - \mu_n)z}. 
\end{equation} 

Thus the paraxial propagation constants are 

\begin{equation} \beta_n^{\mathrm{para}} = n_0 k_0 - \mu_n = n_0 k_0 - \Omega\left(n+\frac12\right). \label{eq:supp-paraxial-beta} 
\end{equation} 

Equation~\eqref{eq:supp-paraxial-beta} shows that the propagation constants themselves are equally spaced: 

\begin{equation} \beta_{n+1}^{\mathrm{para}} - \beta_n^{\mathrm{para}} = -\Omega. 
\end{equation} 

Their spacing is therefore perfectly regular, which is the origin of the exact periodic rephasing in the paraxial model.

\subsection{Exact Scalar Helmholtz Equation}

We now repeat the derivation without the paraxial approximation.

\subsubsection*{Exact Modal Ansatz}

Instead of factoring out a carrier and introducing a slowly varying envelope, we now seek an exact guided mode of the form
\begin{equation}
E(x,z) = \psi(x)\,e^{i\beta z}.
\label{eq:supp-exact-mode-ansatz}
\end{equation}
Inserting Eq.~\eqref{eq:supp-exact-mode-ansatz} into the Helmholtz equation \eqref{eq:supp-helmholtz-1d} yields
\begin{equation}
\frac{d^2\psi}{dx^2}
+
\bigl[k_0^2 n^2(x)-\beta^2\bigr]\psi
= 0.
\label{eq:supp-exact-transverse}
\end{equation}

\subsubsection*{Parabolic Profile in the Exact Equation}

Insert the same parabolic profile as before,
\begin{equation}
n^2(x)=n_0^2-\alpha x^2.
\end{equation}
Then Eq.~\eqref{eq:supp-exact-transverse} becomes
\begin{equation}
\frac{d^2\psi}{dx^2}
+
\bigl[k_0^2 n_0^2-\beta^2-\alpha k_0^2 x^2\bigr]\psi
=0.
\end{equation}
Rearranging,
\begin{equation}
-\frac{d^2\psi}{dx^2}
+
\alpha k_0^2 x^2 \psi
=
\bigl(k_0^2 n_0^2-\beta^2\bigr)\psi.
\label{eq:supp-exact-ho}
\end{equation}
Comparing \eqref{eq:supp-exact-ho} with \eqref{eq:supp-paraxial-ho-rearranged} shows that this again has the form of a harmonic-oscillator eigenvalue problem. The crucial difference from the paraxial case is that the harmonic-oscillator eigenvalue is now
\begin{equation}
\Lambda_n = k_0^2 n_0^2 - \beta_n^2,
\label{eq:supp-Lambda-definition}
\end{equation}
not \(\beta_n\) itself.

Since the harmonic oscillator has equally spaced eigenvalues, one obtains
\begin{equation}
\Lambda_n = (2n+1)\,\Omega',
\qquad n=0,1,2,\dots
\label{eq:supp-Lambda-ladder}
\end{equation}
for some constant \(\Omega'\). Using Eq.~\eqref{eq:supp-Lambda-definition}, this means
\begin{equation}
k_0^2 n_0^2 - \beta_n^2 = (2n+1)\,\Omega',
\end{equation}
or equivalently
\begin{equation}
\beta_n
=
\sqrt{k_0^2 n_0^2-(2n+1)\Omega'}.
\label{eq:supp-exact-beta}
\end{equation}

\subsubsection*{Key Point}

Equation~\eqref{eq:supp-exact-beta} is the central result. It shows that in the exact scalar Helmholtz problem the propagation constants are the square root of an equally spaced sequence. Therefore the \(\beta_n\) themselves are \emph{not} equally spaced.

In other words:
\begin{itemize}
\item In the paraxial approximation, the harmonic-oscillator eigenvalues enter directly into \(\beta_n\), giving an equally spaced \(\beta_n\)-ladder.
\item In the exact scalar Helmholtz equation, the harmonic-oscillator eigenvalues enter into \(k_0^2 n_0^2-\beta_n^2\), so the equally spaced ladder is a ladder in \(\beta_n^2\), not in \(\beta_n\).
\end{itemize}

\subsection{Recovering the Paraxial Limit from the Exact Result}

The paraxial result is recovered by expanding the square root in Eq.~\eqref{eq:supp-exact-beta} when the mode-dependent correction is small:
\begin{equation}
(2n+1)\Omega' \ll k_0^2 n_0^2.
\end{equation}
Then
\begin{align}
\beta_n
&=
\sqrt{k_0^2 n_0^2-(2n+1)\Omega'} \nonumber\\
&\approx
k_0 n_0
-\frac{(2n+1)\Omega'}{2k_0 n_0}.
\label{eq:supp-square-root-expansion}
\end{align}
This is now linear in \(n\), so the propagation constants become approximately equally spaced again. Thus the paraxial harmonic-oscillator picture is simply the first-order expansion of the exact Helmholtz result.

\subsection{Extension to Two Transverse Dimensions}

The two-dimensional transverse problem used in the main text follows exactly the same logic. Starting from
\begin{equation}
\frac{\partial^2 E}{\partial x^2}
+
\frac{\partial^2 E}{\partial y^2}
+
\frac{\partial^2 E}{\partial z^2}
+
k_0^2 n^2(x,y)\,E
=0,
\end{equation}
and using the exact mode ansatz
\begin{equation}
E(x,y,z)=\psi(x,y)e^{i\beta z},
\end{equation}
one obtains
\begin{equation}
\left[
-\frac{\partial^2}{\partial x^2}
-\frac{\partial^2}{\partial y^2}
+
V(x,y)
\right]\psi
=
\bigl(n_1^2 k_0^2 - \beta^2\bigr)\psi,
\end{equation}
where \(V(x,y)\propto x^2+y^2\) for a parabolic GRIN profile. This is the two-dimensional harmonic oscillator. The equally spaced ladder therefore appears in the quantity \(n_1^2 k_0^2-\beta_{mn}^2\), while the exact propagation constants \(\beta_{mn}\) retain a square-root dependence on the total mode order \(m+n+1\). Only after the paraxial expansion does one recover the familiar linear dependence of \(\beta_{mn}\) on \(m+n+1\).

\subsection{Connection to Modal Rephasing}

This distinction is precisely what matters for self-imaging. Exact global rephasing requires that there exist a propagation distance \(z\) and a global phase \(\phi\) such that
\begin{equation}
e^{i\beta_j z}=e^{i\phi}
\qquad \text{for all guided modes } j.
\label{eq:supp-exact-rephasing-condition}
\end{equation}
Equivalently,
\begin{equation}
(\beta_j-\beta_k)z = 2\pi N_{jk},
\qquad N_{jk}\in\mathbb{Z},
\label{eq:supp-exact-rephasing-condition-differences}
\end{equation}
for all pairs of modes \(j\) and \(k\).

In the paraxial model, the \(\beta_j\) are equally spaced, so this condition can be satisfied exactly at a common pitch length. In the exact Helmholtz problem, the square-root dependence destroys this exact regular spacing, so different mode pairs accumulate phase differences at slightly different rates. This is the origin of the residual modal dephasing observed beyond the paraxial approximation.

\subsection{Summary}

The result can be summarized as follows:
\begin{enumerate}
\item A parabolic GRIN profile leads to a harmonic-oscillator problem both in the paraxial approximation and in the exact scalar Helmholtz equation.
\item In the paraxial approximation, the harmonic-oscillator eigenvalues enter directly into the propagation constants \(\beta_n\), leading to an equally spaced \(\beta_n\)-ladder and exact periodic rephasing.
\item In the exact scalar Helmholtz equation, the harmonic-oscillator eigenvalues enter into \(n_0^2 k_0^2-\beta_n^2\), so the equally spaced ladder is a ladder in \(\beta_n^2\), not in \(\beta_n\).
\item The paraxial result is recovered by linearizing the square root in the exact expression.
\item Consequently, exact global modal rephasing is a feature of the paraxial approximation, while the full Helmholtz problem generally supports only approximate rephasing.
\end{enumerate}

This explains why the self-imaging of a parabolic GRIN fiber is exact in the paraxial model but only approximate in the full modal description used for the present MAD-CPA analysis.

\section{Why Optimal Rephasing and Minimum Intermodal Dispersion Yield Nearly the Same Profile Exponent}
\label{sec:supp-rephasing-vs-dmd}

In this section we explain why, for the power-law GRIN family
\begin{equation}
n(r)=n_1\sqrt{1-2\Delta\left(\frac{r}{a}\right)^q},
\qquad 0\le r\le a,
\label{eq:supp-grin-profile}
\end{equation}
the exponent \(q\) that minimizes intermodal dispersion is, to first and second order in \(\Delta\), essentially the same as the exponent that optimizes monochromatic rephasing. The parabolic case \(q=2\) is the classical leading-order reference profile for minimizing modal dispersion in multimode graded-index fibers \cite{Olshansky1976}. We use  conventions 
\((\nu,N,\delta)\) from \cite{Ghatak2004}, where
\begin{equation}
\beta_\nu^2 \simeq k_0^2 n_1^2
\left[
1-2\Delta\left(\frac{\nu}{N}\right)^{\frac{q}{q+2}}
\right],
\qquad
N=\frac{q}{q+2}\,k_0^2 a^2 n_1^2\Delta,
\qquad
k_0=\frac{\omega}{c},
\label{eq:supp-beta-wkb}
\end{equation}
and
\begin{equation}
\delta \equiv \Delta\left(\frac{\nu}{N}\right)^{\frac{q}{q+2}},
\qquad
0<\delta\lesssim \Delta.
\label{eq:supp-delta-def}
\end{equation}
With this notation,
\begin{equation}
\beta_\nu = n_1k_0\sqrt{1-2\delta}.
\label{eq:supp-beta-delta}
\end{equation}

\subsection{Intermodal dispersion}

The modal transit time is
\begin{equation}
t_\nu = L\frac{d\beta_\nu}{d\omega}.
\label{eq:supp-tnu-def}
\end{equation}
Neglecting material dispersion, \(n_1\) and \(\Delta\) are independent of \(\omega\), and since \(N\propto k_0^2\propto \omega^2\), one finds
\begin{equation}
\delta(\omega)\propto \omega^{-p},
\qquad
p\equiv \frac{2q}{q+2}.
\label{eq:supp-p-def}
\end{equation}
Differentiating \eqref{eq:supp-beta-delta} and expanding in powers of \(\delta\) gives the textbook expression
\begin{equation}
t_\nu
=
\frac{n_1L}{c}
\left[
1
+\frac{q-2}{q+2}\,\delta
+\frac{3q-2}{q+2}\,\frac{\delta^2}{2}
+O(\delta^3)
\right].
\label{eq:supp-tnu-expansion}
\end{equation}
Thus the modal delay spread is governed by the \(\delta\)-dependent terms. The minimax criterion used in \cite{Ghatak2004} is to choose \(q\) such that the fastest and slowest guided modes have equal transit times, i.e.
\begin{equation}
t_\nu(\delta=0)=t_\nu(\delta=\Delta).
\label{eq:supp-minimax-condition}
\end{equation}
Using \eqref{eq:supp-tnu-expansion}, this yields
\begin{equation}
\frac{q-2}{q+2}\Delta
+
\frac{3q-2}{q+2}\frac{\Delta^2}{2}
=0,
\end{equation}
and therefore
\begin{equation}
q_0
=
\frac{4+2\Delta}{2+3\Delta}
=
2-2\Delta+O(\Delta^2).
\label{eq:supp-qopt-first}
\end{equation}
If one keeps one more term in the expansion,
\begin{equation}
t_\nu
=
\frac{n_1L}{c}
\left[
1
+\frac{q-2}{q+2}\delta
+\frac{3q-2}{2(q+2)}\delta^2
+\frac{5q-2}{2(q+2)}\delta^3
+O(\delta^4)
\right],
\label{eq:supp-tnu-thirdorder}
\end{equation}
the same minimax procedure gives
\begin{equation}
q_0
=
2-2\Delta-\Delta^2+O(\Delta^3)
\simeq 2\sqrt{1-2\Delta}.
\label{eq:supp-qopt-second}
\end{equation}

\subsection{Monochromatic rephasing criterion}

For monochromatic propagation at fixed \(\omega\), the field can be written as
\begin{equation}
E(z)\propto \sum_\nu a_\nu e^{i\beta_\nu z}.
\end{equation}
Perfect rephasing at a distance \(z=z_p\) would require the propagation constants to form an exact arithmetic ladder in an appropriate mode-group index. For the parabolic case \(q=2\), one obtains the approximate harmonic-oscillator spectrum, and \(\beta_\nu\) is nearly linear in the mode-group label. Beyond the paraxial approximation, however, this linearity is only approximate.

To quantify the deviation from perfect rephasing, it is convenient to introduce a normalized continuous mode coordinate \(x\in[0,1]\) through
\begin{equation}
\delta = \Delta x^p,
\qquad
p=\frac{2q}{q+2}.
\label{eq:supp-delta-x}
\end{equation}
Equivalently,
\begin{equation}
\beta(x)=\beta_0\sqrt{1-2\Delta x^p},
\qquad
\beta_0\equiv n_1k_0.
\label{eq:supp-beta-x}
\end{equation}
For exact self-imaging, \(\beta(x)\) should be as close as possible to a linear function of \(x\). We therefore define the linear interpolation between the endpoints
\begin{equation}
\beta_{\mathrm{lin}}(x)=\beta(0)+x[\beta(1)-\beta(0)],
\label{eq:supp-beta-lin}
\end{equation}
and the residual
\begin{equation}
r(x)=\beta(x)-\beta_{\mathrm{lin}}(x).
\label{eq:supp-rx-def}
\end{equation}
The quantity \(r(x)\) is a direct measure of the deviation from an exact arithmetic ladder and hence of the failure of exact rephasing.

\subsection{Expansion around the parabolic case}

The optimal profile is expected to be close to the parabolic case \(q=2\), so we write
\begin{equation}
p=1+\varepsilon,
\qquad
\varepsilon=O(\Delta).
\label{eq:supp-eps-def}
\end{equation}
The scaling \(\varepsilon=O(\Delta)\) is natural: the intrinsic nonlinearity of the square root already produces corrections of order \(\Delta^2\), and to compensate these by adjusting the exponent one needs \(\Delta\varepsilon\sim \Delta^2\), hence \(\varepsilon\sim \Delta\).

Expanding \eqref{eq:supp-beta-x} for small \(\Delta\), one finds
\begin{equation}
\frac{\beta(x)}{\beta_0}
=
1-\Delta x^p-\frac{\Delta^2}{2}x^{2p}
-\frac{\Delta^3}{2}x^{3p}
+O(\Delta^4).
\label{eq:supp-beta-expanded}
\end{equation}
Now expand for \(p=1+\varepsilon\). To the order needed,
\begin{align}
x^p &= x^{1+\varepsilon}
= x e^{\varepsilon\ln x}
\simeq
x\left(1+\varepsilon\ln x+\frac{\varepsilon^2}{2}(\ln x)^2\right),\\
x^{2p} &= x^{2+2\varepsilon}
\simeq
x^2\left(1+2\varepsilon\ln x\right),\\
x^{3p} &\simeq x^3.
\end{align}
Inserting this into \eqref{eq:supp-beta-expanded} gives
\begin{equation}
\frac{\beta(x)}{\beta_0}
\simeq
1
-\Delta x
-\Delta\varepsilon\,x\ln x
-\frac{\Delta\varepsilon^2}{2}x(\ln x)^2
-\frac{\Delta^2}{2}x^2
-\Delta^2\varepsilon\,x^2\ln x
-\frac{\Delta^3}{2}x^3.
\label{eq:supp-beta-expanded2}
\end{equation}
The term \(-\Delta x\) is the leading linear ladder. The remaining terms are the first deviations from exact rephasing.

\subsection{A simple measure of rephasing quality and the first-order optimum}

A simple criterion for good rephasing is to minimize the curvature of \(\beta(x)\) near the center of the guided spectrum. Indeed, the residual \(r(x)\) vanishes at the endpoints by construction, and the deviation from the chord is controlled by the second derivative of \(\beta(x)\). The midpoint \(x=1/2\) is therefore the most natural location at which to suppress the leading curvature.

Differentiating \eqref{eq:supp-beta-expanded2} twice gives
\begin{equation}
\frac{\beta''(x)}{\beta_0}
\simeq
-\Delta\varepsilon\,\frac{1}{x}
-\Delta\varepsilon^2\,\frac{\ln x+1}{x}
-\Delta^2
-\Delta^2\varepsilon\,(2\ln x+3)
-3\Delta^3 x.
\label{eq:supp-beta-second-derivative}
\end{equation}
Evaluating this at \(x=1/2\), where \(\ln x=-\ln 2\), yields
\begin{equation}
\frac{\beta''(1/2)}{\beta_0}
\simeq
-2\Delta\varepsilon
-2\Delta\varepsilon^2(1-\ln 2)
-\Delta^2
-(3-2\ln 2)\Delta^2\varepsilon
-\frac{3}{2}\Delta^3.
\label{eq:supp-beta-second-mid}
\end{equation}
At leading nontrivial order one keeps only the terms of order \(\Delta^2\), namely
\begin{equation}
\frac{\beta''(1/2)}{\beta_0}
\simeq
-2\Delta\varepsilon-\Delta^2.
\end{equation}
Suppressing this leading curvature gives
\begin{equation}
-2\Delta\varepsilon-\Delta^2=0
\quad\Longrightarrow\quad
\varepsilon=-\frac{\Delta}{2}.
\label{eq:supp-eps-leading}
\end{equation}
Since \(p=1+\varepsilon\), this yields
\begin{equation}
p_{\mathrm{opt}}
=
1-\frac{\Delta}{2}+O(\Delta^2).
\label{eq:supp-popt-leading}
\end{equation}
Converting back to \(q\) using
\begin{equation}
p=\frac{2q}{q+2}
\qquad\Longleftrightarrow\qquad
q=\frac{2p}{2-p},
\label{eq:supp-q-from-p}
\end{equation}
one obtains
\begin{equation}
q_{\mathrm{opt}}
=
2-2\Delta+O(\Delta^2).
\label{eq:supp-qopt-rephasing-leading}
\end{equation}
Comparing this with \eqref{eq:supp-qopt-first} we see that to first order in \(\Delta\), the exponent that minimizes intermodal dispersion and the exponent that suppresses the leading nonlinearity of the monochromatic spectrum are identical. The parabolic case \(q=2\) is the classical leading-order reference profile for minimizing modal dispersion in multimode graded-index fibers \cite{Olshansky1976}.

\subsection{Second-order correction}

To go one step further, write
\begin{equation}
\varepsilon = a\Delta+b\Delta^2+O(\Delta^3).
\label{eq:supp-eps-series}
\end{equation}
From the first-order result we already know
\begin{equation}
a=-\frac{1}{2}.
\end{equation}
Inserting \eqref{eq:supp-eps-series} into \eqref{eq:supp-beta-second-mid} and collecting all terms up to order \(\Delta^3\), one finds
\begin{equation}
-2b
-2(1-\ln 2)a^2
-(3-2\ln 2)a
-\frac{3}{2}
=0.
\end{equation}
With \(a=-1/2\), this gives
\begin{equation}
b=-\frac{1+\ln 2}{4}.
\end{equation}
Hence
\begin{equation}
p_{\mathrm{opt}}
=
1-\frac{\Delta}{2}
-\frac{1+\ln 2}{4}\Delta^2
+O(\Delta^3).
\label{eq:supp-popt-second}
\end{equation}
Using \eqref{eq:supp-q-from-p}, this becomes
\begin{equation}
q_{\mathrm{opt}}
=
2-2\Delta-(\ln 2)\Delta^2+O(\Delta^3).
\label{eq:supp-qopt-rephasing-second}
\end{equation}
This second-order result is not exactly identical to the minimax result \eqref{eq:supp-qopt-second}, but it is extremely close:
\begin{equation}
2-2\Delta-(\ln 2)\Delta^2
\qquad \text{vs.} \qquad
2-2\Delta-\Delta^2.
\end{equation}
The difference appears only in the coefficient of the \(O(\Delta^2)\) term, and is numerically very small for the weak-guidance regime \(\Delta\ll 1\).

\newpage
\subsection{Interpretation}

The reason for the near-coincidence of the two optimizations is that both are governed by the same competition between two effects:

\begin{enumerate}
\item the intrinsic anharmonic correction of the spectrum, represented by the \(-\Delta^2 x^2/2\) term in \eqref{eq:supp-beta-expanded2}, and

\item the tunable correction induced by choosing \(q\neq 2\), represented by the \(-\Delta\varepsilon\,x\ln x\) term.
\end{enumerate}

In the group-delay problem these same two effects appear in the \(\delta\)- and \(\delta^2\)-terms of \eqref{eq:supp-tnu-expansion}; in the rephasing problem they appear as the first two deviations from a perfectly linear propagation-constant ladder. Since both optimizations are controlled by the same perturbative balance, they produce the same first-order shift
\begin{equation}
q_{\mathrm{opt}} = 2-2\Delta+O(\Delta^2),
\end{equation}
and only differ very slightly in the coefficient of the second-order term.

\subsection{Conclusion}

Within the weak-guidance WKB description, the exponent that minimizes intermodal dispersion and the exponent that optimizes monochromatic rephasing are nearly identical because both are determined by the same leading non-parabolic correction to the spectrum. The textbook minimax calculation gives
\begin{equation}
q_{\mathrm{opt}}=2-2\Delta-\Delta^2+O(\Delta^3),
\end{equation}
while the present rephasing-based curvature criterion gives
\begin{equation}
q_{\mathrm{opt}}=2-2\Delta-(\ln 2)\Delta^2+O(\Delta^3).
\end{equation}
Thus the two criteria coincide exactly to first order and remain extremely close at second order. This explains why in numerical simulations the optimum for minimal intermodal dispersion and the optimum for best monochromatic rephasing are practically indistinguishable for realistic values of \(\Delta\).

\end{document}